\documentclass[11pt]{article} 
\usepackage[utf8]{inputenc}
\usepackage[english]{babel}
\usepackage{amsmath}
\usepackage[a4paper,margin=1in]{geometry} 
\usepackage{authblk} 
\usepackage[colorlinks=true, linkcolor=black, citecolor=black, urlcolor=blue, filecolor=blue,breaklinks=true]{hyperref} 
\usepackage{setspace} 
\usepackage{csquotes} 
\usepackage{ragged2e} 
\usepackage[capitalise, noabbrev, nameinlink]{cleveref}
\crefdefaultlabelformat{#2\textbf{#1}#3} 
\Crefname{figure}{\textbf{Figure}}{\textbf{Figures}} 
\Crefname{section}{\textbf{Section}}{\textbf{Sections}} 
\Crefname{table}{\textbf{Table}}{\textbf{Tables}} 

\usepackage{graphicx}
\graphicspath{{./main_figs/}{./supplement/suppl_figs/}} 
\DeclareGraphicsExtensions{.pdf,.jpeg,.JPG,.png,.PNG, .eps, .tiff}
\usepackage{subcaption} 
\DeclareCaptionLabelFormat{bold}{\textbf{(#2)}} 
\usepackage{pgffor} 
\usepackage{alphalph} 
\usepackage[labelfont=bf, textfont=bf, singlelinecheck=off, textfont=footnotesize]{caption} 
\usepackage{booktabs}
\usepackage{multirow}
\usepackage{rotating}

\usepackage[dvipsnames]{xcolor} 
\usepackage{changepage} 
\usepackage{enumitem} 
\usepackage{longtable}
\usepackage{array}     
\usepackage{ragged2e}  

\usepackage[style=apa, backend=biber]{biblatex}
\title{Scientific tools and Innovation: Big Science Facilities Yield More Novel and Interdisciplinary Knowledge}

\author[1,2]{Mingze Zhang}
\author[1,2]{Yizhan Li}
\author[3]{Yutong Li}
\author[1,2,*]{Zexia Li}
\affil[1]{National Science Library, Chinese Academy of Sciences, Beijing, China}
\affil[2]{Department of Information Resource Management, University of Chinese Academy of Sciences, Beijing China}
\affil[3]{Ningbo Institute of Materials Technology and Engineering, Chinese Academy of Sciences, Zhejiang, China}

\affil[ ]{ } 
\affil[*]{\textit{Correspondence to }\underline{lizexia@mail.las.ac.cn}}
\date{} 

\newcommand{\makeAbstract}{
\begin{abstract}
Scientific tools dictate the boundaries of human knowledge, serving as the foundation for perceptions and explorations. In the era of Big Science, science are increasingly dependent on advanced analytical technologies and experimental platforms. Over the past decades, national and supranational entities have invested massive financial resources, collaborative networks, and collective intelligence to construct Big Science Facilities (BSFs) aimed at generating cutting-edge knowledge. However, empirical evaluations of these machines' actual performance in driving scientific innovation remain scarce. To address this gap, we collected 310,086 publications from 88 global BSFs and constructed a matched control dataset of approximately 3 million publications sharing the same last authors. Our analysis reveals that the utilization of BSFs has expanded significantly since 1950s. Crucially, publications supported by these facilities exhibit higher recombinant novelty and interdisciplinary integration. Furthermore, this improvement is most pronounced in non-physical sciences—domains traditionally peripheral to BSFs' core focus—indicating the emergence of a powerful intra-facility knowledge spillover effect. By enriching the Facilitymetrics framework, our findings provide empirical evidence that BSFs act as vital engines for scientific discovery, offering policymakers essential metrics to justify infrastructural investments, while prompting the science of science community to reassess the profound impact of scientific tools on knowledge production.
\end{abstract}
}

\begin{document}
\setstretch{1.15} 
    
\maketitle
\makeAbstract
\clearpage 
\section{Introduction}

Historically, paradigm-shifting scientific breakthroughs have rarely stemmed from isolated "eureka" moments of individual genius; rather, they are predominantly catalyzed by the evolution of observational and experimental tools \parencite{hallonsten2016big, borner2021visualizing, price1963little}. In essence, scientific tools fundamentally dictate the epistemic boundaries of what researchers can perceive and explore. Since the mid-20th century, the scale and complexity of these tools have experienced exponential growth, ushering in the "Big Science" era \parencite{hallonsten2017collaborative, lauto2013arge, heinze2017reinvention}. Today, Big Science Facilities (BSFs)—such as synchrotron radiation light sources, particle accelerators, neutron sources, and free-electron laser—have evolved from mere physical tools into complex socio-technical systems \parencite{qiao2016scientific, zhang2021knowledge}. As the products of massive public investment, these facilities inherently shoulder the mission of pushing scientific frontiers and fostering radical innovation \parencite{zhang2025impact, ZHANG2026, hallonsten2013introducing}.

Unlike the closed, military-oriented projects of the past, modern BSFs operate primarily as "user-oriented," open-access infrastructures. Under this "big machine and small team" paradigm \parencite{hallonsten2016big}, external scientists from diverse institutions and disciplines converge to compete for highly limited experimental beamtime \parencite{yang2024beamtimes, soderstrom2023global}. Theoretically, by aggregating diverse intellectual resources, BSFs function as massive "boundary objects" that naturally facilitate cross-disciplinary knowledge recombination and spillover \parencite{zhang2021knowledge, d2019research, soderstrom2022generic, li2025can}. However, a critical question remains: does this extreme concentration of rare resources actually translate into paper-level scientific novelty and interdisciplinarity? Given the exorbitant costs associated with constructing and operating BSFs \parencite{hallonsten2014expensive, zhang2025impact}, policymakers, funding agencies, and the public are increasingly concerned about the Return on Investment (ROI) of these scientific behemoths \parencite{hallonsten2016use}. Therefore, a systematic, quantitative evaluation of the extent to which BSFs substantively advance knowledge production is urgently needed.

In recent years, the emergence of "Facilitymetrics" has provided a valuable lens for assessing the scientific performance of large research infrastructures \parencite{hallonsten2013introducing}. Nevertheless, empirical evaluations in this domain face two major challenges. First, there is a severe data bottleneck. Due to inconsistent database structures and varying levels of data openness across global facilities, existing empirical studies are often constrained to single-case analyses or limited samples of fewer than five facilities \parencite{zhang2025impact, soderstrom2023global, yang2024beamtimes}. Such localized paradigms struggle to quantify the holistic innovation effects of BSFs within the global knowledge production network \parencite{ZHANG2026, lauto2013arge}. Second, previous evaluations often suffer from selection bias and endogeneity. They only focus on those publications within the facilities while access to BSFs is a highly competitive resource disproportionately granted to elite scientists with established reputations \parencite{d2019research}. Consequently, extant studies fail to disentangle the innate innovative capacity of the scientists from the marginal epistemic benefits provided by the tools, potentially overestimating the actual catalytic effect of the facilities. Further, previous studies paid much attentions on evaluating those research performance under the context of big science facilities \parencite{zhang2024facilities, zhang2025impact, ZHANG2026, hallonsten2014expensive, silva2019co, soderstrom2023structure}, focusing on understanding "How to make use of these machines?" but naturally overlooked the exact innovation effects of them and recusal the question "If without these machines, how innovation might be damaged?"

To bridge these critical research gaps, this study constructs a comprehensive, global-scale dataset by manually exporting and computationally extracting publication records from the official websites of 88 BSFs worldwide. By triangulating these records with the OpenAlex bibliographic database, we established a robust dataset comprising approximately three million publications spanning several decades with the last authors (typically representing principal investigators or team leaders) of these publications fixed \parencite{xu2024impact, xu2022flat, lin2023remote, xu2022team}. Methodologically, last author as the fixed-effect could also address the partly endogeneity issue and introduce strict condition into our estimation models \parencite{liu2022further, peng2024author}. This rigorous specification allows us to compare the epistemic outputs of the same scientist and his/her teams when utilizing versus not utilizing big science facilities, thereby isolating the true marginal contribution of the facility from the scientist's baseline capabilities. We then comprehensively quantify the positive relationships of BSFs on scientific performance using atypical journal combinations for novelty \parencite{uzzi2013atypical} and the Rao-Stirling diversity index for interdisciplinary \parencite{wang2015interdisciplinarity, yang2025understanding}.

Our empirical results demonstrate that the utilization of BSFs significantly enhances the probability of producing novel and highly interdisciplinary knowledge. More importantly, our heterogeneity analysis reveals a highly counter-intuitive phenomenon: although BSFs are predominantly designed for core disciplines (mainly physical sciences), they confer disproportionately higher marginal performance benefits to peripheral disciplines (e.g., life, health, and social sciences). This suggests that while mainstream physics may be experiencing diminishing marginal returns along a mature technological trajectory, researchers in secondary domains are harvesting the early dividends of "technology borrowing" and executing "distal interdisciplinary"\parencite{bloom2020ideas, chu2021slowed}. By uncovering these dynamics, this study not only enriches the theoretical and empirical frameworks of Facilitymetrics and Science of Science but also provides robust, data-driven insights for global stakeholders aiming to optimize resource allocation and maximize the innovation potential of big science facilities.

\section{Literature review}

\subsection{Relationship between scientific tools and innovation}
Rather than stemming from isolated "eureka moments" of individual genius, scientific innovations and novel knowledge discoveries are actually benefited from evolution of observational tools. In essence, scientific tools fundamentally dictate the epistemic boundaries of what scientists can perceive and explore. From a historical perspective, the trajectory of scientific advancement has been inextricably linked to the evolution of observational and experimental tools. By pushing the limits of observable phenomena, the deployment of novel scientific facilities directly expands the "adjacent possible, enabling researchers to explore previously inaccessible conceptual spaces.
This tool-driven view of scientific progress is deeply rooted in the foundational theories of the Science of Science and Science and Technology Studies. For instance, Derek de Solla Price fundamentally challenged the traditional linear model of innovation, arguing that the development of new laboratory tools and techniques—frequently precede and dictate theoretical formulations \parencite{price1963little}, rather than the other way around. Modern scientometric studies provide robust empirical evidence echoing this perspective. Recent quantitative analyses have consistently demonstrated that research portfolios supported by novel equipment or specialized infrastructure funding exhibit a significantly higher propensity for generating disruptive and high-impact discoveries\parencite{zhang2025impact, soderstrom2023global}. In this context, the scientific tools are not merely a passive conduit for data collection, but an active participant that reshapes the cognitive and methodological landscape of a discipline \parencite{stephan2017sectoral}.

However, the paradigm of scientific instrumentation has undergone a radical transformation since the mid-20th century, transitioning from isolated "table-top" tools to Big Science Facilities \parencite{stix2001little}. In the era of Big Science, advanced tools such as synchrotron radiation light sources, particle accelerators, and massive astronomical observatories have evolved into complex socio-technical systems \parencite{hallonsten2016big, hallonsten2017collaborative, heinze2017reinvention,manganote2016effect}. This exponential leap in scale fundamentally alters the dynamics of knowledge production. While traditional tools might expand the "adjacent possible" for a single laboratory, Big Science Facilities function as massive "boundary objects" \parencite{star1989institutional}. By requiring immense investment and offering multifaceted technical capabilities \cite{ZHANG2026}, these facilities physically and intellectually aggregate scientists across diverse domains—from physics and materials science to biology and health sciences \parencite{soderstrom2022generic, silva2019co}. Consequently, Big Science Facilities act as structural catalysts, breaking down institutional silos and forcing unprecedented interdisciplinary knowledge recombination, which naturally yields higher epistemic novelty \parencite{d2019research, hinnant2012author, yang2024beamtimes}.

\subsection{Big science era with advanced tools}
\textcite{price1963little} demonstrates that scientific research gradually entered a era of big science during and after the world war II. Since that time, requirements on advanced analytical technologies are increasing day by day \parencite{kozheurov2020megascience, heinze2017reinvention}. Currently, big science has already become one of the most feature of modern science \parencite{crease2016new} and amounts of big science projects are launched by national or supranational governments around the global for the past several decades \parencite{borner2021visualizing}, with huge, expensive, and complex facilities constructed \parencite{heidler2015qualifying,hallonsten2014expensive}. These facilities, known as big science facilities, are reported to become the engine of scientific discoveries, particularly in the STEM-related domains \parencite{borner2021visualizing,manganote2016effect}. Currently, it is reported that Nobel Prize show close associations with big science facilities \parencite{stix2001little}. For instance, the experimental confirmation of the Higgs boson relied heavily on the Large Hadron Collider at CERN \parencite{brumfiel2011collider}, and the first direct observation of gravitational waves was made possible by the highly complex LIGO observatories \parencite{castelvecchi2017gravitational}.
There is no doubt that big science facilities strongly push the limits of human knowledge, placing the academia on an endless journey to chase the frontiers of science \parencite{hallonsten2017collaborative}. 

Currently, most big science facilities around the globe follow a "user-oriented" pattern \parencite{d2019research}. Consequently, big science has developed a new feature: research at these facilities is typically conducted by small groups of external users rather than massive, centralized scientific teams \parencite{englert2023hyperauthorship}. This paradigm is defined as the "big machine and small team" model by \parencite{hallonsten2016big}. These external users—a cohort of scientists predominantly affiliated with academia and rarely from industry—submit research proposals to the facilities they intend to use, awaiting approval and the assignment of experimental beamtime \parencite{yang2024beamtimes}. Under this context, big science facilities serve as massive experimental infrastructures equipped with cutting-edge technologies, offering advanced capabilities for new knowledge discovery \parencite{hallonsten2017collaborative, heinze2017reinvention}. However, they represent an exceedingly rare resource for scientific research, as their annual operational capacities are strictly limited \parencite{bianco2017waypoints, hallonsten2016use}, which triggers fierce intellectual competition among all potential users \parencite{linovel}. To mitigate risks, the assignment of these limited resources tends to prioritize external users with strong academic backgrounds and high reputations, thereby providing assurances of beneficial scientific returns to the facilities \parencite{lauto2013arge, kozheurov2020megascience}. However, while prioritizing elite researchers might theoretically maximize the probability of success, it remains to be proven whether this intensive concentration of rare resources actually yields the anticipated breakthroughs \parencite{giffoni2023public}.

\subsection{Essential evaluations on big science facilities' scientific performance and facilitymetrics}

As above mentioned, big science facilities are huge infrastructures with expensive public investments. They are also associated with the nature of scientific discoveries and knowledge advancements \parencite{hallonsten2016big}. To response the concerns from the public, governments, funders, and scientists, it is quite important to understand the concrete effects of big science facilities' benefits on scientific development and social progress \parencite{giffoni2023public}. In short, all stakeholders are question about the return of investments (ROI) of big science facilities \parencite{hallonsten2014expensive}. Therefore, about past a decade, \textcite{hallonsten2013introducing} proposed Facilitymetric, which applied methodologies from scientometrics to evaluate the scientific performance of big science facility. It has become a systematic research orientation that unique perspective is discussed, customised indicators are constructed, and the research context of big science facilities still remain lots of research gap to be filled up.

In respect to the evaluations of scientific performance, facility intense index and facility impact factor are proposed \parencite{hallonsten2016use}, which both attempt to avoid the statistical misrepresentation in ROI caused by the simple counts of scientific productivity or impacts \parencite{hallonsten2014expensive}. One study, based on the publications supported by big science facilities, evaluate the level of interdisciplinary in the context of big science facilities research \parencite{soderstrom2022generic}. Further study visualized the service range of big science facilities around the globe, demonstrating the intersected natures of global share and regional strengths of the facilities \parencite{soderstrom2023global}. 

More importantly, using these publications yielded from big science facilities, previous work compared the different of topic selections between teams only with external scientists (known as users) and teams also included internal scientists (a group of scientists affiliated to the facilities) \parencite{silva2019co}. Such classification of teams uncover an important factor, the effect of internal scientists, since under the context of big science facilities, internal scientists with unique knowledge background that could support users to make use of these advanced technologies \parencite{soderstrom2023structure}. The optimized usages of facilities might drive to produce more scientific ideas and are unconsciously fell into the research area of team science \parencite{wong2014team}. Extant studies not only qualitatively discussed the interaction mechanism \parencite{d2019research} between internal and external scientists but also compare the gaps in the scientific performance between two type of teams. Results show that teams with internal scientists are associated with higher novelty, scientific disruption, and impacts \parencite{zhang2025impact, zhang2024facilities}. Such evaluations response to the strategies of make use of big science facilities.

A recent study depicts a novel type of collaboration, supporting one publication by more than one big science facility, and defines such case as "co-utilization" \parencite{linovel, ZHANG2026}, which also contribute to understand the global relationships and collaborative innovation within these facilities themselves and knowledge flows originated from global users \parencite{soderstrom2023global}. Additionally, the evaluation of big science facilities' performance also show valuable perspective for the framework of science of science. Given that most users have to use the facility on-site, one study demonstrates that those users, supported by more than one facility in one year, could be applied to quantify the effects of short-term scientific travel on scientific performance \parencite{zhangscientific}. 

However, the above-mentioned scientific performance evaluations based on or focus on big science facilities discussed the significances of scientific tools only in the theory and limited, since they only focus on the inside of big science facility context rather than providing enough, overall, and empirical-based evidence to ensure how these facilities contribute to knowledge advancements and novel knowledge discoveries by comparing with the condition that we do not have them. 
Therefore, a clear research gap emerged and is more crucial response to stakeholders' concerns by answering the exact innovation effects from these expensive facilities instead of merely demonstrating that how to make use of them.

Our study aims to provide comprehensive empirical evidence and contribute to bridge this knowledge gap. One inescapable reason for previous studies remain this gap might because of the complicated process of data collection. Global facilities possess different constructions of literature databases and diverse open level of their publication data (See details in subsection 3.1 Data collection). The challenges in data collection drive extant studies only focus on no more than five facilities around the globe \parencite{zhang2025impact}, leading to a research manner similar to the case study. Therefore, such settings deeply undermined the possibilities to quantify globally innovation effects of these expensive facilities.

We collected the about 88 main facilities, with different types and diverse functions, around the world to overcome this data bottleneck, thereby shifting the research paradigm from localized case studies to a truly global and systematic evaluation. Under this context, we could truly demonstrate how the big science facilities are associated with the production of cutting-edge, novel, advanced, and integrated knowledge. Furthermore, our research design provides insights on understanding the benefits of advanced scientific tools.

\clearpage 
\section{Data \& Method}

\subsection{Data collection}
To answer the research questions we proposed above-mentioned, the collection of publications supported by big science facilities is the initial step. Following the framework of Facilitymetrics \parencite{zhang2025impact, silva2019co, soderstrom2023global}, we collected the BSFs' publications from the publication lists or self-constructed databases from their own official websites rather than retrieve from commonly used bibliographical databases such as Web of Science or Scopus. Since most big science facilities are public-invested infrastructures, it is their responsibility to display their achievements to the public and taxpayers. Therefore, facilities always collate their publications and available them on their websites \parencite{ZHANG2026}. Moreover, publications from commonly used databases might suffer data loss and cause extreme bias, Noted that the identifications are always based on acknowledgment texts and affiliation addresses. However, not all papers acknowledge the facilities in a standard way and even some papers do not mention the facilities their used \parencite{soderstrom2023global}. Additionally, using affiliation may only include those publications with internal scientists participation but overlook the main part of facilities outputs \parencite{d2019research, soderstrom2023structure}. 

Therefore, in May 2025, we collected 88 big science facilities' publication from the globe by considering the level and quality of openness. We also referred to the recommendations from expertise in Chinese Academy of Sciences, ensuring that collection includes the most and mainstream user-oriented facilities around the world (see Supplementary Table S1 in supplementary materials for the detailed list of facilities). Notably, given that the publication data from the diverse sources and different facilities open inconsistent metadata, it is essential for us to introduce large-scale bibliographical database as supplementary data to break the limits of websites' data \parencite{priem2022openalex}. We therefore applied OpenAlex to finish our research design, and it is known as a fully open access database for science of science that has been used widely across the global community \parencite{liu2025ai, xu2024impact, yang2025repeat, peng2024author}. Our version of OpenAlex is updated in Jan. 2025.

After matching with OpenAlex, we ensured that 359,301 published records yielded by 88 facilities but there are only 310,086 publications, which informs that the existence of co-utilization according to our previous work\parencite{ZHANG2026, linovel}. Noted that we only consider English publication and confine the document type by article for high accuracy \parencite{yang2025quantifying, zhang2024impact}.

To understand the performance variations between BSF and Non-BSF publications, we collected Non-BSF publications based on the last authors of BSF publications. Specifically, in our BSF datasets, 310,086 publications are authored by 82,438 authors in the last position (including the single authored publications). We further retrieve these authors' publications in OpenAlex and construct an expanded dataset to determine the innovation effects of big science facilities. Eventually, we access a dataset including 3,644,181 publications, 7.99\% pulications are supported by BSF, by 80,480 last authors, from 1950 to 2025 (See Figure \ref{fig:fig1}a for annual distributions.).

\subsection{Scientific Novelty and Interdisciplinary}
We applied atypical combinations of journals to quantify the scientific novelty, which proposed by \textcite{uzzi2013atypical} in 2013 and is reproduced by amounts of previous work in the domain of science of science \parencite{ke2026geography, wagner2019international, tian2025distinctive, yang2022gender}. This metric, rooted in Schumpeterian innovation theory\parencite{li2025can, malerba1995schumpeterian}, defines that atypical combinations of prior knowledge mean novelty while common combinations demonstrate the conventionality of scientific papers and the combinations are represented by co-cited journals. The differences, captured by the z-score, between observed frequency and the expected frequency for every journal combinations, normalized by the standard errors according to the stimulated co-citation network, demonstrate whether the combination is uncommon or not \parencite{uzzi2013atypical}. For every publication with a few journal references, a z-score distribution could be accessed and 10\textsuperscript{th} z-score is considered to represent the novelty score (NS) of the publication. Following the previous work \parencite{yang2022gender, zhang2025impact, ke2026geography}, a novel publication is defined as a publication with NS lower than zero and we therefore adopted the method and access the original novelty score of every publication in our dataset from SciSciNet-v2 \parencite{lin2023sciscinet}.

In respect to the interdisciplinary, the extant indicators measure various perspectives of interdisciplinary, including variety, balance, and disparity \parencite{yang2025unraveling, yang2025understanding, wang2015interdisciplinarity}. All of them are suitable according to the definition of interdisciplinary \parencite{newell2001theory}. However, we adopt Rao-stirling to measure the interdisciplinary because it is a further comprehensive indicator \parencite{leydesdorff2019interdisciplinarity}, considering the three perspectives together and is widely acknowledged by previous works \parencite{zhang2016diversity, li2026interdisciplinary}. The formula of Rao-stirling is as follows:

\begin{equation}
Rao-stirling_{i,t} = \sum_{m,n \in cf_{i}} d_{m,n,t}\, R_{m}\, R_{n}
\end{equation}

Given that paper \textit{i}, published in \textit{t} year and from its references, the paper's citing fields (\textit{cf}\textsubscript{i}) represented by \textit{m} and \textit{n} and \textit{d}\textsubscript{m,n,t} records the distance between field \textit{m} and field \textit{n} in \textit{t} year measured by cosine similarity based on the all publications' fields in the \textit{t} year. Further, \textit{R}\textsubscript{m} and \textit{R}\textsubscript{n} represent the ratio of fields \textit{m} and \textit{n} cited by the focal paper \textit{i}. The Rao-stirling of every publication is calculated by the authors based on the citation relationships and 252 subfields provided by OpenAlex (version: Jan 2025).

\subsection{Big Science Facility Usage}
The identification of big science facility usage constitute the main independent variable in this study and according to the details in the section of data collection. We considered the publications collected from big science facilities' websites as publications used big science facilities. Further, according to the our previous work \parencite{zhang2025impact, ZHANG2026}, the existences of co-utilizations between or among big science facilities enable us to consider the number of facilities used as an alternative independent variable for robustness check.

\subsection{Control Variables}
According to the previous literature, we control several confounding factors to avoid potential influence on the production of novel and interdisciplinary knowledge.
Several factors are controlled: (1) the number of authors; (2) the number of institutions; (3) the number of countries; These three factors are considered as the effects of teams on scientific innovation are widely justified despite the diverse type of teams, for instance, inter-institutional or international teams \parencite{yoo2024interaction, lin2023remote, thelwall2023coauthored}, gender or racial-diverse teams \parencite{zhang2024impact, gonzalez2025woman, yang2022gender, peng2024author}, and teams with authors from different academic level and different background \parencite{xu2022team, xu2024impact}. (4) the number of references: Notably, the volume of references show significant relationship with the number of journal combinations and the volume of citing fields \parencite{zhang2016diversity, uzzi2013atypical}. (5) the global composition of team leadership: we consider the leadership, the combination of the first and the last author, following the common definition in science of science \parencite{xu2024impact, xu2022flat, liu2022further}. Therefore, there are three types of team leadership: Both North, Both South, and North-South Combination, and we further take North-South Combination as reference group in the regression. (6) the average career age: according to the previous work \parencite{xu2022team}, the average career age represents the power level of teams and leads a close relationship with scientific performance. (7) the average institution prestige: we applied h-index to capture the institution prestige and the level of h-index implies that the teams power in scientific research contributed by their affiliation \parencite{li2025gender, peng2025gender}. (8) the journal h-index: different level of journal also show diverse preference on novelty and interdisciplinary \parencite{liang2023bias, rafols2012journal}, and therefore we also adopted h-index to measure the academic level of journals. (9) the Core journal: based on the journal classification provided by CWTS \parencite{van_eck_2024_10949671}, we classify all journals in our dataset into Core journal and non-Core journal. 

As for the fixed effects, we consider the discipline, adopted from 26 fields provided by OpenAlex, and publication year. Additionally, last author is considered as another fixed effect, which enables us to estimate the effects of BSFs on novel and interdisciplinary knowledge production within individual authors.

\subsection{Summary and Estimated Model}

The summary of regression variables and fixed-effects could be seen in Table \ref{tab:var_summary}, where we summarize the involved variable and provide a basic description.

\begin{table}[htbp]
    \centering
    \caption{Summary and description of regression variables and fixed-effects}
    \small
    \setlength{\tabcolsep}{3pt}
    \begin{tabular}{c|c|p{7cm}}
    \hline\hline
    Variable types & Variable names& Description \\\hline
       Dependent Var.  & $Novelty_{i,a,d,t}$ & A binary variable determines that whether the publication yield atypical combinations or not. \\
         & $Interdisciplinary_{i,a,d,t}$ & A continuous variable, quantifying the interdisciplinary level of the focal publication.\\\hline
       Independent Var.  & $BSF\ pub_{i,a,d,t}$ & A binary variable determines that whether the publication is supported by big science facilities or not.\\\hline
       Control Var.  & $Number\ of\ authors\ (log)_{i,a,d,t}$  & A continuous variable, quantifying the number of authors in the focal publication.\\
         & $Number of institutions\ (log)_{i,a,d,t}$ & A continuous variable, quantifying the number of institutions in the focal publication.\\
         & $Number\ of\ countries\ (log)_{i,a,d,t}$ & A continuous variable, quantifying the number of countries in the focal publication.\\
         & $Number\ of\ references\ (log)_{i,a,d,t}$ & A continuous variable, quantifying the number of references in the focal publication.\\
         & $Leadership(Global\ types)_{i,a,d,t}$ & A categorical variable that determines the composition of the focal publication's leadership. (Mixed, all North, and all South).\\
         & $Avg.\ career age\ (log)_{i,a,d,t}$& A continuous variable, quantifying the average career age of all co-authors in the focal publication.\\
         & $Avg.\ institution\ h-index\ (log)_{i,a,d,t}$& A continuous variable, quantifying the average institution prestige of all co-authors in the focal publication. \\
         & $Journal\ h-index\ (log)_{i,a,d,t}$& A continuous variable, quantifying the level of the published journal of the focal publication.\\
         & $Core\ Journal_{i,a,d,t}$  & A binary variable that determines whether the published journal of the focal publication is a core journal or not.\\\hline
         Fixed-effects& Publication year & The published year of the focal publication.\\
         & Research disciplines & The research fields (26 in total) of the focal publication\\
         & Last author & The last author of the focal publication\\ 
    \hline\hline     
    \end{tabular}
    \label{tab:var_summary}
\end{table}

We apply logistic regression on novelty metric since we transformed the novelty score into a binary variable following the widely accepted method from previous work \parencite{yang2022gender, zhang2025impact, wagner2019international}. We conduct ordinary least squares regression on interdisciplinary ability given that rao-stirling is a continuous metric roughly distributed from 0 to 0.4. In both models, we incorporate fixed effects for disciplines, publication year, and the last author, thereby absorbing unobserved confounding variations across these dimensions. See the following regression formula (2).
\begin{equation}
DV_{i,a,d,t} = \beta_0 + \beta_1 \mathit{BSF pub}_{i,a,d,t} + \gamma Z_{i,a,d,t} + \delta X_{i,a,d,t} + A_a + D_d + Y_t + \varepsilon_{i,a,d,t}
\end{equation}
Where $DV_{i,a,d,t}$ includes $Novelty_{i,a,d,t}$ and $Interdisciplinary_{i,a,d,t}$, indicating the knowledge performance of the focal publication \textit{i} from research disciplines \textit{d}, authored by last author \textit{a} in \textit{t} year. The $BSF\ pub_{i,a,d,t}$ demonstrate that whether the focal publication \textit{i} is supported by big science facilities or not. $Z_{i,a,d,t}$ denotes a set of control variables at paper-level while $X_{i,a,d,t}$ is a set of journal-level controls. Paper-level controls include $Number\ of\ authors\ (log)_{i,a,d,t}$, $Number of institutions\ (log)_{i,a,d,t}$, $Number\ of\ countries\ (log)_{i,a,d,t}$, $Number\ of\ references\ (log)_{i,a,d,t}$,  $Leadership(Global\ types)_{i,a,d,t}$, $Avg.\ career age\ (log)_{i,a,d,t}$, and $Avg.\ institution\ h-index\ (log)_{i,a,d,t}$. Journal-level controls is consist of $Journal\ h-index\ (log)_{i,a,d,t}$ and $Core\ Journal_{i,a,d,t}$. All specifications include last author ($A_a$), research disciplines ($D_d$), and publication year ($Y_t$) fixed-effects.
\clearpage 
\section{Results}

\subsection*{Increasing usages of big science facilities}
\begin{figure}[htbp]
    \centering
    \includegraphics[width=1\linewidth]{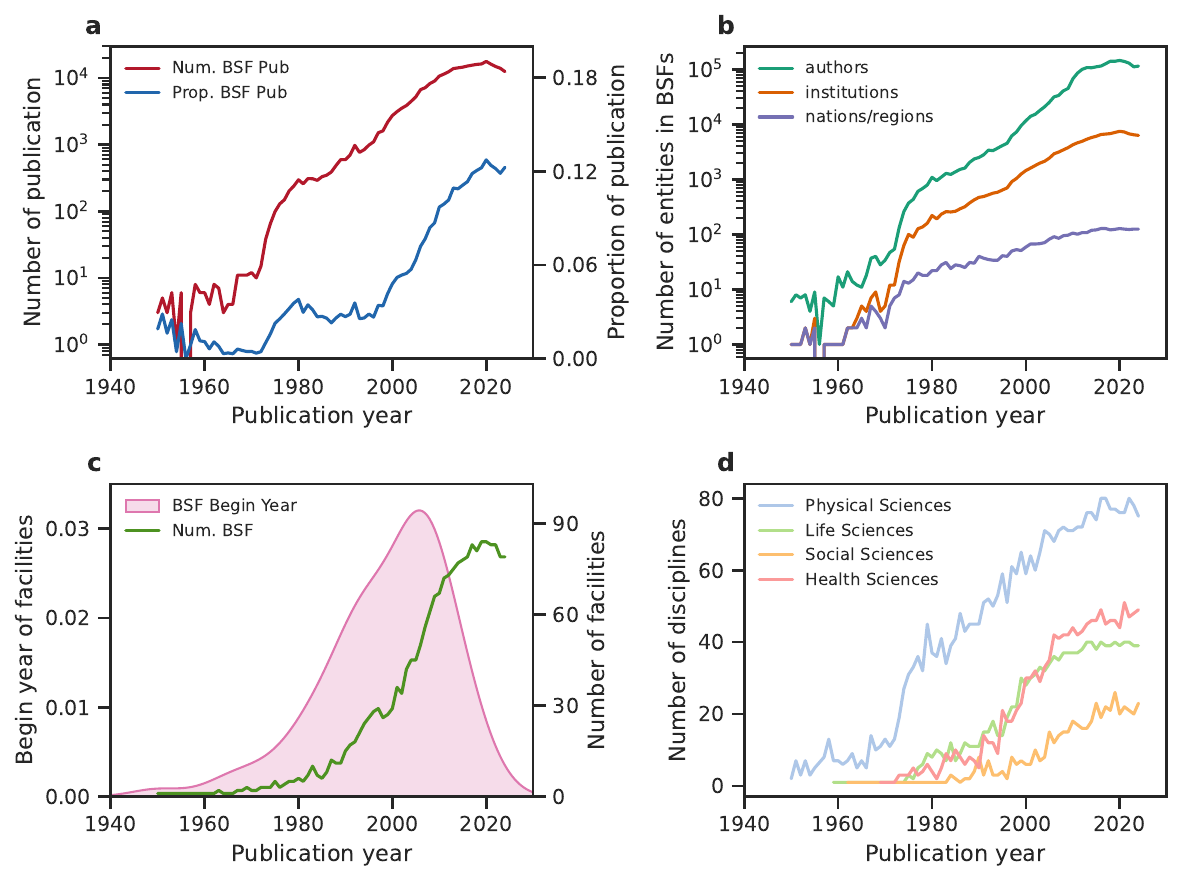}
    \caption{\textbf{Basic descriptions demonstrate that utilization of big science facilities would be future trend.} \textbf{a,} The increasing trends (numbers in red and proportions in blue) in publications supported by big science facility from 1950s to 2020s. \textbf{b,} The increasing trends of authors, institutions, and nations in using big science facilities for scientific publications. \textbf{c,} The numbers of big science facilities (in green) from 1950s to 2020s and the distributions of their begin year (in pink). \textbf{d,} Disciplines in four research areas all witness the increasing trends in using big science facilities (blue for physical sciences, green and pink for life and health sciences respectively, and orange for social sciences.)}
    \label{fig:fig1}
\end{figure}

From Figure \ref{fig:fig1}a, we could observe the increasing trend in using big science facilities in the scientific research since the annual publication increased to more than 10,000 papers in recent years and in 1980s, only 100 papers are published annually (see red line). The proportions of publications supported by big science facilities (compare with our constructed dataset by focusing on the last authors) increase from about 0.03 to 0.12 about past four decades (see blue line). In Figure \ref{fig:fig1}b, we visualize the annual number of authors, institutions, and nations/regions using big science facilities. The number of authors (in green) increased to 100,000 annually at the beginning of 2020s while there are about 10,000 institutions (in orange) around the globe yearly use the facilities and about 100 nations/regions (in purple) are supported by big science facilities. In Figure \ref{fig:fig1}c, we show the distribution of the begin years of 88 facilities (see pink area) in our collected data set and the annual number of facilities with yielding the scientific publication (see green line). It seems that around the beginning of $21_{st}$ century, the construction of big science facilities reach to the peak while at that time there are only about 30 out of 88 facilities are involved in scientific research. However, at the end of 2020, almost all facilities in our collected data set are observed with publications. In Figure \ref{fig:fig1}d, we count the number of disciplines (255 in total, provided by OpenAlex) involved in the research context of big science facilities across four main research domains. We could tell that the big science facilities become increasing interdisciplinary. Research disciplines from physical sciences show the greatest preference to using big science facilities that there are already more than 70 disciplines are involved. The number of disciplines in health sciences and life sciences are similar and around 2020, about 40 disciplines are associated with big science facilities. The number of disciplines decreases to about 20 in the domain of social sciences around 2020.

\subsection*{The benefits of big science facilities on Novelty and Interdisciplinary}

\begin{figure}[htbp]
    \centering
    \includegraphics[width=1\linewidth]{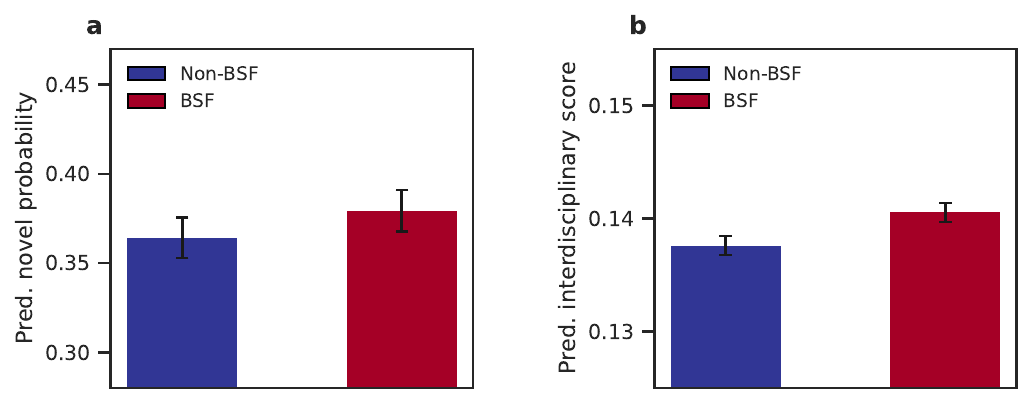}
    \caption{\textbf{Papers supported by big science facilities show higher novel probability and interdisciplinary ability.} \textbf{a,} Papers supported by BSF associated with 37.92\% (95\%CI=[36.29\%, 39.56\%]) novel probability while papers are not supported by BSF show 36.41\%(95\%CI=[34.80\%, 38.02\%]) novel probability \textbf{b,} Papers supported by BSF show 0.141 (95\%CI=[0.139, 0.142])  while those papers without supports from BSF show 0.138 (95\%CI=[0.136, 0.139]) abilities in interdisciplinary. Error bars represent 95\% confidence intervals based on standard errors.}
    \label{fig:fig2}
\end{figure}

In Figure \ref{fig:fig2}, we visualize the predicted results of using (in red) and not using (in blue) big science facilities on producing novel (see Figure \ref{fig:fig2}a) and interdisciplinary knowledge (see Figure \ref{fig:fig2}b). The corresponding regression models could be seen in Table \ref{tab:reg}. The performance gaps are significant even we add controls and fixed-effects with potential impacts. According to Figure \ref{fig:fig2}a, papers supported by big science facilities show 37.92\% probability to produce novel knowledge in average while papers are not supported by big science facilities only with 36.41\% novel probability even if they are originated from the same last author (1.04 times higher in relative). According to the interdisciplinary, in Figure \ref{fig:fig2}b, big science facilities supported papers are averagely 0.141, relatively 1.02 times higher than papers without supports of big science facilities within the same last author.

In Table \ref{tab:reg}, we display the final regression models of novelty and interdisciplinary for simplicity and the step-wise regression could be seen in Supplementary Table S2 and Supplementary Table S3 respectively.
From Table \ref{tab:reg}, the coefficients of producing novel and interdisciplinary knowledge by big science facilities is 0.084 ($p<0.01$) and 0.003 ($p<0.001$) respectively, providing positive evidence of better scientific performance for big science facilities knowledge production. Three fixed-effects, research disciplines, publication year, and the last authors, are all considered, demonstrating the reliability of our discoveries. In regression model of novelty, 2,682,339 papers are considered and the pseudo R$^2$ is 0.186. 2,768,157 papers are involved in the regression of interdisciplinary with 0.514 R$^2$

\begin{table}[htbp]
\caption{\textbf{Fixed-effects regressions determine the positive effects of big science facilities on novel and interdisciplinary knowledge production.} In column (1), we display the regression result on novelty and in column (2), result on interdisciplinary is shown. Both models are considered paper-level and journal-level controls with three fixed-effects to avoid unobservable factors. Standard-errors are shown in parentheses and the significant Codes: ***: p$<$0.01, **: p$<$0.05, *: p$<$0.1.}
\label{tab:reg}
\begingroup \centering
\footnotesize
\renewcommand{\arraystretch}{0.5}
\begin{tabular}{lcc}
\tabularnewline \midrule \midrule 
& Novel paper & Interdisciplinary\\
& (1) & (2)\\
\midrule
BSF pub = True & 0.084$^{***}$ & 0.003$^{***}$\\   
               & (0.006) & (0.000)\\                  
Number of authors (log) & 0.257$^{***}$ & $5.38\times 10^{-5}$\\ 
                        & (0.005) & (0.000)\\
Number of institutions (log) & 0.010 & 0.001$^{***}$\\ 
                             & (0.006) & (0.000)\\ 
Number of countries (log)  & -0.282$^{***}$  & -0.003$^{***}$\\   
                           & (0.009)         & (0.000)\\    
Number of references (log) & 0.155$^{***}$ & 0.002$^{***}$\\   
                           & (0.003)       & (0.000)\\              
Leadership = all North & 0.133$^{***}$ & 0.001$^{***}$\\   
                       & (0.009)       & (0.000)\\
Leadership = all South & -0.015 & 0.0008$^{*}$\\   
                       & (0.014)& (0.000)\\
Avg. career age (log) & 0.032$^{***}$ & 0.0006$^{***}$\\   
                      & (0.005)       & (0.000)\\
Avg. institution h-index (log)  & -0.025$^{***}$ & 0.0004$^{***}$\\   
                                & (0.004)        & (0.000)\\ 
Journal h-index (log) & -0.045$^{***}$ & -0.0008$^{***}$\\   
                      & (0.002)        & (0.000)\\
Core Journal = True & 0.082$^{***}$ & $4.81\times 10^{-5}$\\    
                    & (0.012)       & (0.000)\\
\midrule
\emph{Fixed-effects controls}\\
Discipline (26 fields) & Yes & Yes\\
Publication year & Yes & Yes\\
Last Author & Yes & Yes\\
\midrule 
Observations & 2,682,339 & 2,768,157\\
(Pseudo) R$^2$ & 0.186 & 0.514\\
\midrule \midrule\\
\end{tabular} \par\endgroup
\end{table}

\subsection*{The heterogeneous effects of big science facilities on Novelty and Interdisciplinary}

\begin{figure}[htbp]
    \centering
    \includegraphics[width=1\linewidth]{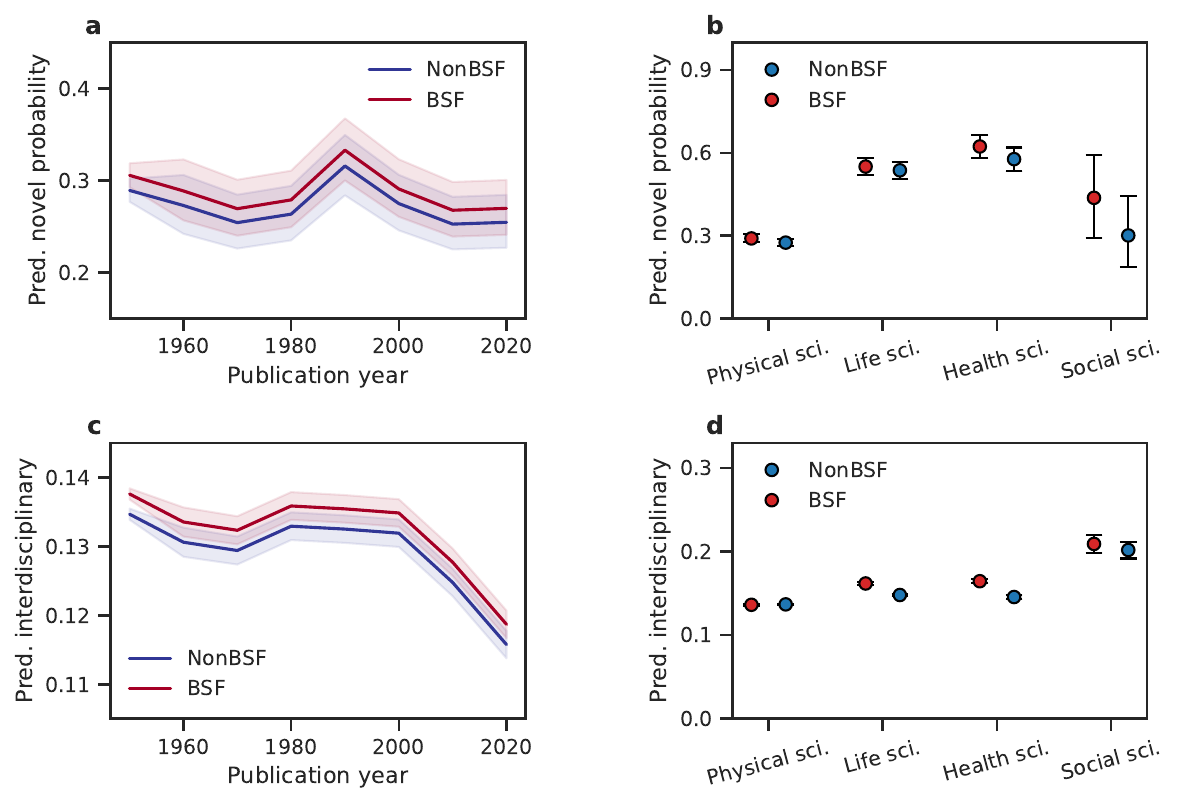}
    \caption{\textbf{Big science facilities show consistent effects on producing novel and interdisciplinary knowledge.} \textbf{a,} Predicted probability of novel knowledge between BSF and Non-BSF supported publications annually. \textbf{b,} Predicted probability of novel knowledge within four research areas. We fit a separate model for each broad area (N = 2,236,396 for Physical science; N = 683,985 for Life science; N = 337,939 for Health science; N = 26,000 for Social science). See details of regression in Supplementary Table S4. \textbf{c,} Predicted level of interdisciplinary knowledge between BSF and Non-BSF supported publications. \textbf{d,} Predicted level of interdisciplinary knowledge within four research areas. We fit a separate model for each broad area (N = 2,321,151 for Physical science; N = 729,686 for Life science; N = 376,452 for Health science; N = 41,437 for Social science). See details of regression in Supplementary Table S5. Error bars represent 95\% confidence intervals based on standard errors.}
    \label{fig:fig3}
\end{figure}

According to Figure \ref{fig:fig3}, we conduct heterogeneous analysis to estimate the effects of big science facilities on contributing to novel and interdisciplinary knowledge production.
In Figure \ref{fig:fig3}a and Figure \ref{fig:fig3}b, we visualize the regression results about novelty within different decades and four main research domains while the same results about interdisciplinary are shown in Figure \ref{fig:fig3}c and Figure \ref{fig:fig3}d.

Across almost 7 decades (from 1950s to 2020s) in our dataset, big science facilities show stable contributions to producing novel and interdiscipinary knowledge. According to the marginal effects predicting from regression models, in 1950s (see Figure \ref{fig:fig3}a), publications supported by big science facilities are associated with 30.6\% probability (95\%CI = [29.3\%, 31.9\%]) to be novel in average and at the same, publications without big science facilities' supports only show averagely 28.9\% probability (95\%CI = [27.7\%, 30.2\%]). Around 2020s, big science facilities' publications show about 27.0\% novel probability (95\%CI = [24.1\%, 30.1\%]), relatively 1.06 times higher than publications without using big science facilities (25.5\%, 95\%CI = [22.7\%, 28.4\%]) in spite of they all are authored by the same last author.
Similar trend could be observed in interdisciplinary ability between publications with/without big science facilities (Figure \ref{fig:fig3}c). At the beginning of our observed years (1950s), big science facilities' publications show 0.1376 (95\%CI = [0.1368, 0.1384]) averagely in interdisciplinary score while during 2020s, big science facilities still possess 0.1187 (95\%CI = [0.1167, 0.1207]). All observed decades demonstrate that big science facilities assist those last authors to produce about 1.03 times interdisciplinary ability higher than the publications when they are not assisted by the facilities (in 1950s, 0.1347 on average with 95\%CI = [0.1338, 0.1355]; in 2020s, 0.1158 on average with 95\%CI = [0.1138, 0.1178]).

In respect to four main research domains, big science facilities keep advantages in producing novel knowledge (see Figure \ref{fig:fig3}b for visualization and Supplementary Table S4 for regression results) while are not useful to produce interdisciplinary knowledge particularly in disciplines related to physical sciences (see Figure \ref{fig:fig3}d for visualization and Supplementary Table S5 for regression results). 

According to producing novel knowledge within specific research domains, big science facilities show coefficients about 0.590 (p$<$0.01, N = 337,939) in social sciences and 0.191 (p$<$0.01, N = 26,000) in health sciences (see columns (3) and (4) in Supplementary Table S4). As a result, the marginal predictions (see Figure \ref{fig:fig3}b) report that publications with big science facilities supports associated with averagely 62.3\% (95\%CI = [58.1\%, 66.4\%]) and 43.7\% (95\%CI = [53.4\%, 62.0\%]) probabilities to produce novel knowledge in health and social sciences respectively. For those publications without big science facilities in health and social sciences authored by the same last authors, the novel probabilities are averagely 57.7\% (95\%CI = [53.4\%, 62.0\%]) and 30.1\% (95\%CI = [18.8\%, 44.4\%]). In research disciplines related to life sciences, publications supported by big science facilities show about 55.0\% (95\%CI = [51.9\%, 58.2\%]) probability to novelty, which is 1.02 times relatively higher to the publications within the same research domain and authored by the same last authors (about 53.7\%, 95\%CI = [50.5\%, 56.8\%]).

Big science facilities supported publications could show higher interdisciplinary level except for physical sciences (see Figure \ref{fig:fig3}d and Supplementary Table S5). Especially, when focusing on research disciplines related to the life sciences and health sciences, big science facilities positively associated with higher interdisciplinary  knowledge since the coefficients about 0.014 and 0.019 (both p$<$0.01, N = 729,686 and 376,452 respectively). However, publications supported by big science facilities show significantly less ability to interdisciplinary in physical science ($\beta = -0.001$, p$<$0.01). In the corresponding marginal results, big science facilities contribute about 1.09 and 1.13 relative increase in life sciences (0.1618, 95\%CI = [0.1601, 0.1635] vs 0.1481, 95\% = [0.1464, 0.1497]) and health sciences (0.1645, 95\%CI = [0.1620, 0.1671] vs 0.1456, 95\% = [0.1431, 0.1482]).

Actually, research disciplines related to physical sciences are the major research area for big science facilities (N = 2,236,396 and 2,321,151 for novelty and interdisciplinary respectively). However, big science facilities show only 29.0\% (95\%CI = [27.6\%, 30.4\%]) probability to novelty and 0.1363 (95\%CI = [0.1355, 0.1372]) in interdisciplinary ability (lower than without big science facilities, where with 0.1368 on average, 95\%CI = [0.1360, 0.1377]). These heterogeneous effects across four research domains imply minor domains are more benefit and suggest that the knowledge spillover might be occurred in the context of big science facilities.

\subsection*{Robustness checks}
To validate our discoveries, we firstly exclude the last author fixed-effect to estimate the overall effects of big science facilities on producing novel (see Table \ref{tab:rob} column (1)) and interdisciplinary ((see Table \ref{tab:rob} column (2))) knowledge. Without authors fixed, publications supported by big science facilities show positive association with novelty ($\beta$ = 0.107, p$<$0.01) and interdisciplinary ($\beta$ = 0.005, p$<$0.01). Comparing with the regression results with authors fixed, we could not only observe the robust significances but also improved effect sizes. This indicates that the effect is driven not only by the within-author variations strictly isolated in our main analysis, but also by the broader cross-sectional differences between authors. The consistency across both specifications strongly supports the robustness of the identified mechanisms that big science facilities supported science could yield more novel and interdisciplinary knowledge.
According to the columns (3) and (4) of Table \ref{tab:rob}, we apply a continuous independent variable to validate the significantly positive relationships between using big science facilities and producing novel ($beta$ = 0.063, p$<$0.01) and interdisciplinary knowledge ($beta$ = 0.002, p$<$0.01) under the contexts of three fixed-effects. Specifically, the variable, the number of facilities, is measured by how many facilities are used to support the focal publication and according to our previous work, it is not a rare sense and using more than one facilities could be defined as "Co-utilization" \parencite{ZHANG2026}. 
In the columns (5) and (6) of Table \ref{tab:rob}, we adopted new novelty metrics from \parencite{arts2025beyond}, which measures the scientific novelty based on natural language processing (NLP) and proposes the creations of new words and new phrases of each publications in the OpenAlex from 1900-2023. We therefore transform the number of new words and new phrases into binary variables to determine whether the focal publication in our dataset with or without new creation. Therefore, we also apply fixed-effects logistic regressions to estimate the positive effects of big science facilities on creating novel knowledge. Results validate the robustness of our main discoveries (New words: $\beta$ = 0.050, p$<$0.01; New phrases: $\beta$ = 0.027, p$<$0.01).

\begin{table}[htbp]
\centering
\caption{\textbf{Robustness checks ensure the effects of BSFs on novel and interdisciplinary knowledge production.} In columns (1) and (2), we display the positive effects of big science facilities on producing novel and interdisciplinary knowledge without considering the last author fixed-effect; In columns (3) and (4), we replace the binary independent variable by a continuous variable, which demonstrates the number of facilities used by the focal publication, to estimate the robustness of main discoveries; In columns (5) and (6), we adopted a new metric about novelty from \parencite{arts2025beyond}. Specifically, we measure the abilities of big science facilities' publications to create new words and new phrases under the context of three fixed-effects. Standard-errors are shown in parentheses and the significant Codes: ***: p$<$0.01, **: p$<$0.05, *: p$<$0.1.}
\label{tab:rob}
\begingroup \centering
\footnotesize
\renewcommand{\arraystretch}{0.5}
\setlength{\tabcolsep}{1pt}
\begin{tabular}{lcccccc}
\tabularnewline \midrule \midrule 
& Novel paper & Interdisciplinary & Novel paper & Interdisciplinary & New words & New phrases\\
& (1) & (2) & (3) & (4) & (5) & (6)\\
\midrule
BSF pub = True & 0.107$^{***}$ & 0.005$^{***}$ & & & 0.050$^{***}$& 0.027$^{***}$\\   
               & (0.004) & (0.000) & & & (0.012)& (0.007)\\
Number of facilities &  & & 0.063$^{***}$ & 0.002$^{***}$\\   
               &  & & (0.005)& (0.000) \\   
Number of authors (log) & 0.121$^{***}$ & -0.005$^{***}$ & 0.257$^{***}$& $5.4\times 10^{-5}$ & 0.304$^{***}$& 0.248$^{***}$\\ 
                        & (0.003) & (0.000) & (0.005)& (0.000)& (0.009)& (0.005)\\
Number of institutions (log) & -0.024$^{***}$ & 0.002$^{***}$ & 0.010 & 0.001$^{***}$ & -0.065$^{***}$& -0.049$^{***}$\\ 
                        & (0.004) & (0.000) & (0.006)& (0.000)& (0.011)& (0.007)\\ 
Number of countries (log)  & -0.208$^{***}$  & $9.92\times 10^{-5}$ & -0.283$^{***}$& -0.003$^{***}$& 0.035$^{**}$& -0.008\\   
                           & (0.007)         & (0.000) & (0.009)& (0.000)& (0.017)& (0.010)\\    
Number of references (log) & 0.095$^{***}$ & 0.003$^{***}$ & 0.155$^{***}$& 0.002$^{***}$& -0.021$^{***}$& 0.068$^{***}$\\   
                           & (0.002)       & (0.000) & (0.004)& ($6.48\times 10^{-5}$) & (0.003)& (0.002)\\              
Leadership = all North & 0.085$^{***}$ & 0.002$^{***}$ & 0.133$^{***}$& 0.001$^{***}$& 0.001& 0.026$^{**}$\\   
                       & (0.007)       & (0.000) & (0.009)& (0.000)& (0.019)& (0.011)\\
Leadership = all South & -0.213$^{***}$ & -0.003$^{***}$ & -0.016& 0.001$^{*}$& 0.011& 0.018\\   
                       & (0.008)& (0.000) & (0.014)& (0.000)& (0.029) & (0.017)\\
Avg. career age (log) & -0.033$^{**}$ & 0.0001 & 0.032$^{***}$& 0.001$^{***}$& -0.048$^{***}$& -0.062$^{***}$\\   
                      & (0.003)       & (0.000) & (0.005)& (0.000)& (0.009)& (0.006)\\
Avg. institution h-index (log)  & -0.007 & 0.001$^{***}$ & -0.024$^{***}$& 0.000$^{***}$& -0.020$^{***}$& 0.010$^{**}$\\   
                                & (0.002)        & (0.000) & (0.004)& (0.000)& (0.007)& (0.004)\\ 
Journal h-index (log) & -0.119$^{***}$  & -0.003$^{***}$ & -0.044$^{***}$& -0.001$^{***}$& 0.057$^{***}$& 0.107$^{***}$\\   
                      & (0.002)        & (0.000) & (0.002)& (0.000)& (0.004)& (0.003)\\
Core Journal = True & 0.183$^{***}$ & 0.001$^{**}$ & 0.083$^{***}$& $5.58\times 10^{-5}$& 0.796$^{***}$& 0.734$^{***}$\\    
                    & (0.010)       & (0.000) & (0.012)& (0.000)& (0.018)& (0.011)\\
\midrule
\emph{Fixed-effects controls}\\
Discipline (26 fields) & Yes & Yes& Yes & Yes& Yes & Yes\\
Publication year & Yes & Yes& Yes & Yes& Yes & Yes\\
Last Author & No & No& Yes & Yes& Yes & Yes\\
\midrule 
Observations & 2,760,381 & 2,768,157 & 2,682,339 & 2,768,157&1,620,451&1,831,114\\
(Pseudo) R$^2$ & 0.044 & 0.222 & 0.186 & 0.514 & 0.109 & 0.123\\
\midrule \midrule
\end{tabular} \par\endgroup
\end{table}

\clearpage

\section{Discussion and Conclusions}
Building on self collecting publication data from about 90 worldwide big science facilities, this study matched a data sample by focusing on the last authors to demonstrate how these scientific tools under the big science era contribute to novel and interdisciplinary knowledge discoveries. According to extant knowledge and viewpoints about innovative effects of scientific tools \parencite{soderstrom2023global, price1963little}, the positive outcomes might be expected, especially for those involved scientists. However, for other stakeholders in the community, for instance, funding grants, policy-makers, and the public, this study provides strong empirical evidence to ensure the positive effect of these huge machines and demonstrates one of the necessities of constructing these machines \parencite{hallonsten2014expensive, hallonsten2013introducing}. 

Specifically, we collected 310,086 publications, authored by 82,438 last authors and supported by 88 big science facilities, and matched about 3 millions publications to set up a comparable dataset. Based on the dataset, this study firstly ensures that big science facilities significantly contribute to novel and interdisciplinary knowledge discoveries, which enriches the research frameworks and theories of Facilitymetrics and provides a novel insight for global scientometrics community to evaluate the effects of scientific tools. 
The innovation effects of these big science facilities are significantly positive even if we considered the last authors as fixed effects. In the research domain of Science of Science, the last authors are usually taken as the corresponding author, principal investigator, and authors in the leadership \parencite{xu2022flat, xu2024impact}. Therefore, control them could be a useful proxy of same research team and put two groups of publications under the same research quality tiers.
Additionally, under the current research systems, these facilities are mostly constructed for physical related research disciplines while they show less applicable abilities for experiments related to life, health, and social sciences \parencite{hallonsten2016big, soderstrom2022generic, silva2019co}. However, the estimated coefficients observed that big science facilities tend to benefit those peripheral disciplines more rather than core disciplines. This differential impact fundamentally captures a profound knowledge spillover effect of these huge machines. From the perspective of recombinant innovation \parencite{uzzi2013atypical, wagner2019international, malerba1995schumpeterian}, importing advanced instrumental capabilities from physics into biology or health sciences creates atypical knowledge combinations, thereby generating substantial novelty. Furthermore, while physics may be experiencing diminishing marginal epistemic returns from these established tools, peripheral disciplines are harvesting the early dividends of technology borrowing, leading to disproportionately higher interdisciplinary and novelty yields \parencite{bloom2020ideas,chu2021slowed, jones2009burden}.

Our results in the subsection of robustness check further ensure the positive effects of big science facilities. If we omit the author fixed effect, the estimated coefficients of these huge machines on producing novel and interdisciplinary knowledge could be substantially amplified. Further, we considered the co-utilizations of big science facilities \parencite{ZHANG2026} and ensure using more advanced scientific tools are more beneficial to novel and interdisciplinary knowledge discoveries. By adopting text metrics, we ensure the abilities to create new concepts for the scientific communities of big science facilities, which stably validates our main discoveries.

Despite its contributions, this study has several limitations that present valuable avenues for future research. First, because our dataset relies on publications manually exported and computationally extracted from the websites of 88 large-scale scientific facilities, our sample is inevitably constrained by the accessibility and varying quality of these public repositories \parencite{ZHANG2026, soderstrom2023structure, silva2019co, soderstrom2023global} and some facilities are therefore not included in our sample. Given that some degree of data loss and sample omission is unavoidable, we advocate for a higher degree of openness in facility-level publication data to facilitate more rigorous scientific performance evaluations \parencite{d2019research, soderstrom2022generic}. 
Further, this study primarily establishes a correlational, rather than causal, relationship between the use of Big Science facilities and the production of novel, interdisciplinary knowledge. Because access to these facilities is a highly competitive and scarce resource, it is highly plausible that the awarded projects inherently possess higher baseline levels of novelty and interdisciplinarity—a classic selection bias. To address this endogeneity issue, we suggest that future research leverage public data on facility application proposals. By comparing the subsequent publication outcomes of successful versus unsuccessful applicants, future studies could employ quasi-experimental designs to more accurately isolate the true causal impact of these scientific tools.
Moreover, although OpenAlex has become a widely adopted bibliographic database within the Science of Science community, inherent data quality concerns cannot be overlooked. Specifically, OpenAlex has known coverage limitations regarding institutional affiliations \parencite{zhang2024missing}, reference linkages \parencite{culbert2025reference}, and journal metadata \parencite{zheng2025understanding}, which could potentially introduce bias into our findings. We encourage future studies to triangulate open-source data with established proprietary databases (e.g., Web of Science, Scopus) or specialized institutional repositories to achieve more robust insights into the innovation dynamics of these facilities.
Additionally, constrained by the scope of this study, our analysis focuses exclusively on journal articles to proxy the knowledge production of authors with and without access to large-scale scientific facilities. However, we acknowledge that the epistemic outputs of these massive scientific tools are not confined to traditional journal publications \parencite{yang2024beamtimes, zhang2021knowledge, borner2021visualizing}. Their broader impacts encompass doctoral dissertations, technical reports, open scientific datasets, and science popularization efforts \parencite{bianco2017waypoints, heinze2017reinvention}. These diverse artifacts yield long-term benefits for academia and the broader innovation ecosystem \parencite{li2022managing, qiao2016scientific}, extending far beyond the immediate generation of novel and interdisciplinary knowledge. Therefore, we encourage future research to evaluate these facilities through a holistic, systems-level perspective, comprehensively assessing their multifaceted benefits across science, society, and the economy.

Eventually, by directing the global academic community's attention toward the epistemic role of scientific tools in the Big Science era, this study serves as a foundational step. We highly encourage future research to further enrich the theoretical and empirical frameworks of Facility-metrics, thereby providing more enlightening discoveries and actionable insights for all stakeholders.

\section*{Acknowledgements}
This work was supported by the National Social Science Fund Major Projects of China (Project No.
22\&ZD127). We would like to thank Hao PENG and Zhesi SHEN for their insightful discussion and valuable comments from reviewers.

\section*{Declaration of Competing Interest}
The authors declare that they have no competing financial interests or personal relationships that could have influenced the work reported in this paper.

\section*{Data Availability}
The datasets generated during and/or analysed during the current study are available in the Mingze ZHANG repository, \url{https://github.com/zhangmingze-ss/BSF_innovation}. The data of OpenAlex could be accessed at: \url{https://docs.openalex.org/}.

\begin{singlespace}
\printbibliography
\end{singlespace}
\clearpage

\end{document}


\setcounter{table}{0}
\renewcommand{\thetable}{S\arabic{table}}

\setcounter{figure}{0}
\renewcommand{\thefigure}{S\arabic{figure}}

\begin{center}
    \Large Supplementary materials for \\[0.5em]
    \Large \textbf{Scientific tools and Innovation: Big Science Facilities Yield More Novel and Interdisciplinary Knowledge} \\[0.5em]
    \large by Mingze Zhang, Yizhan Li, Yutong Li, and Zexia Li. \\[0.5em]
    \large \today
\end{center}

\begin{spacing}{1.0}
\footnotesize
\renewcommand{\arraystretch}{0.85}
\setlength{\tabcolsep}{3pt}
\begin{longtable}[c]{>{\RaggedRight\arraybackslash}m{5.5cm}|cccccc}
\caption{\textbf{Details of self-collected publication data from 88 global big science facilities.} These facilities are ranked by country alpha-2 code (ISO 3166) in the Alphabetical order. Notably, in column of \textit{Type}, SLS: Synchrotron Light Source; NS: Neutron Source; MS: Muon Source; MAG: High Magnetic Field Facilities; FEL: Free-Electron Laser; NO: Neutrino Observatory; NC: Other types.}
\label{tab:facility}
\\\hline\hline
\textbf{Facility name}	&	\textbf{Type}	& \textbf{Country}	&	\textbf{Abbr.}	& \textbf{Num. pub}	& \textbf{Begin year}	& \textbf{End year}\\\hline
Center for the Advancement of Natural Discoveries using Light Emission	&	SLS	&	AM	&	CANDL	&	122	&	2013	&	2024	\\
National Deuteration Facility	&	NC	&	AU	&	NDF	&	172	&	2011	&	2024	\\
Australian Centre for Neutron Scattering	&	NS	&	AU	&	ACNS	&	2,007	&	2007	&	2024	\\
Australian Synchrotron	&	SLS	&	AU	&	AS	&	5,280	&	2008	&	2025	\\
Laboratório Nacional de Luz Síncrotron	&	SLS	&	BR	&	LNLS	&	4,736	&	1987	&	2022	\\
SNO lab	&	NO	&	CA	&	SNO	&	107	&	2008	&	2022	\\
Canadian Institute for Neutron Scattering	&	NS	&	CA	&	CINS	&	1,626	&	1950	&	2019	\\
Canadian Light Source	&	SLS	&	CA	&	CLS	&	3,358	&	2003	&	2024	\\
Swiss X-ray Free Electron Laser	&	FEL	&	CH	&	SFEL	&	167	&	2006	&	2024	\\
Swiss Muon Source	&	MS	&	CH	&	SMS	&	1,379	&	1991	&	2024	\\
Swiss Spallation Neutron Source	&	NS	&	CH	&	SINQ	&	2,233	&	1996	&	2024	\\
Swiss Research InfraStructure for Particle physics	&	PP	&	CH	&	CHRISP	&	1,581	&	1990	&	2024	\\
Swiss Light Source	&	SLS	&	CH	&	SLS	&	9,139	&	1999	&	2024	\\
Aarhus Storage Ring in Denmark	&	SLS	&	DK	&	ASTRID	&	1,006	&	1983	&	2024	\\
ALBA	&	SLS	&	ES	&	ALBA	&	2,697	&	1996	&	2024	\\
European XFEL	&	FEL	&	EU	&	EXFEL	&	884	&	2007	&	2024	\\
European Magnetic Field Laboratory	&	MAG	&	EU	&	EMFL	&	2,073	&	2003	&	2024	\\
Extreme Light Infrastructure	&	NC	&	EU	&	ELI	&	1,096	&	2011	&	2024	\\
Institut Laue–Langevin	&	NS	&	EU	&	ILL	&	14,430	&	1963	&	2024	\\
European Synchrotron Radiation Facility	&	SLS	&	EU	&	ESRF	&	28,984	&	1987	&	2024	\\
Grand National Heavy Ion Accelerator	&	PP	&	FR	&	GANIL	&	1,202	&	2011	&	2024	\\
Soleil	&	SLS	&	FR	&	SOLEIL	&	4,976	&	1997	&	2024	\\
Free-Electron LASer in Hamburg	&	FEL	&	GE	&	FLASH	&	201	&	2,007	&	2024	\\
Free-Electron LASer in Hamburg 2	&	FEL	&	GE	&	FLASH2	&	42	&	2017	&	2024	\\
DESY NanoLab	&	NC	&	GE	&	DESYNL	&	55	&	2015	&	2024	\\
High-Power Radiation Sources	&	NC	&	GE	&	ELBE	&	265	&	2004	&	2024	\\
Electron Microscopy Core Facility	&	NC	&	GE	&	EMCF	&	117	&	2009	&	2024	\\
German Engineering Materials Science Centre	&	NC	&	GE	&	GEMS	&	328	&	2007	&	2024	\\
Ion Beam Centre	&	NC	&	GE	&	IBC	&	3,425	&	1989	&	2024	\\
PHELIX	&	NC	&	GE	&	PHELIX	&	487	&	1996	&	2024	\\
BER II	&	NS	&	GE	&	BER II	&	1,277	&	1992	&	2024	\\
Forschungs-Neutronenquelle Heinz Maier-Leibnitz	&	NS	&	GE	&	FRM II	&	1,385	&	1994	&	2024	\\
Jülich Centre for Neutron Science	&	NS	&	GE	&	JCNS	&	1,142	&	2007	&	2024	\\
DORIS III	&	PP	&	GE	&	DORIS III	&	4,022	&	1979	&	2023	\\
Universal Linear Accelerator	&	PP	&	GE	&	UNILAC	&	77	&	2001	&	2024	\\
Berlin Electron Storage Ring Society for Synchrotron Radiation II	&	SLS	&	GE	&	BESSY II	&	6,292	&	2003	&	2024	\\
PETRA III	&	SLS	&	GE	&	PETRA III	&	5,059	&	1995	&	2024	\\
Free Electron Laser Radiation for Multidisciplinary Investigations	&	FEL	&	IT	&	FERMI	&	373	&	2004	&	2024	\\
Elettra Sincorotrone Trieste	&	SLS	&	IT	&	ELETTRA	&	6,793	&	1994	&	2024	\\
SACLA	&	FEL	&	JP	&	SACLA	&	273	&	2012	&	2024	\\
J-PARC MLF-MS	&	MS	&	JP	&	MLF-MS	&	132	&	2009	&	2024	\\
J-PARC MLF-NS	&	NS	&	JP	&	MLF-NS	&	1,329	&	2006	&	2024	\\
Photon Factory	&	SLS	&	JP	&	PF	&	13,534	&	1969	&	2024	\\
SPring-8	&	SLS	&	JP	&	SPring-8	&	13,019	&	2005	&	2024	\\
PAL-XFEL Facility	&	FEL	&	KR	&	PAL	&	45	&	2017	&	2024	\\
Pohang Light Source I	&	SLS	&	KR	&	PLS-I	&	2,368	&	1996	&	2014	\\
Pohang Light Source II	&	SLS	&	KR	&	PLS-II	&	6,546	&	2008	&	2024	\\
Solaris	&	SLS	&	PL	&	SOLARIS	&	149	&	2019	&	2024	\\
MAX-IV, Accelerators-FEL	&	FEL	&	SE	&	MAX-IV-FEL	&	112	&	1982	&	2024	\\
MAX-IV	&	SLS	&	SE	&	MAX-IV	&	2,591	&	1993	&	2024	\\
Singapore Synchrotron Light Source	&	SLS	&	SG	&	SSLS	&	185	&	1995	&	2024	\\
ISIS	&	NS	&	UK	&	ISIS	&	13,610	&	1975	&	2024	\\
Accelerators and Lasers In Combined Experiments	&	PP	&	UK	&	ALICE	&	383	&	2012	&	2024	\\
Compact Linear Accelerator for Research and Applications	&	PP	&	UK	&	CLARA	&	25	&	2017	&	2024	\\
Central Laser Facility 	&	PP	&	UK	&	CLF	&	2,530	&	1967	&	2024	\\
Muon Ionisation Cooling Experiment	&	PP	&	UK	&	MICE	&	4	&	2012	&	2021	\\
Linac Coherent Light Sourece	&	FEL	&	US	&	LCLS	&	5	&	2011	&	2023	\\
National High Magnetic Field Laboratory	&	MAG	&	US	&	NHMEL	&	7,812	&	1980	&	2024	\\
Advanced Modular Incoherent Scatter Radar	&	NC	&	US	&	AMISR	&	284	&	2004	&	2024	\\
Atmospheric Radiation Measurement	&	NC	&	US	&	ARM	&	4,355	&	1986	&	2024	\\
Biomass Feedstock National User Facility	&	NC	&	US	&	BFNUF	&	92	&	2006	&	2020	\\
Center for Functional Nanomaterials	&	NC	&	US	&	CFN	&	1,872	&	2007	&	2023	\\
Center for Integrated Nanotechnologies	&	NC	&	US	&	CINT	&	2,977	&	2003	&	2024	\\
Center for Nanoscale Materials	&	NC	&	US	&	CNM	&	3,268	&	2001	&	2024	\\
Center for Nanophase Materials Sciences	&	NC	&	US	&	CNMS	&	1,498	&	2002	&	2024	\\
Environmental Molecular Sciences Laboratory	&	NC	&	US	&	EMSL	&	3,680	&	2003	&	2022	\\
Joint Genome Institute	&	NC	&	US	&	JGI	&	2,866	&	2007	&	2024	\\
National Spherical Torus Experiment – Upgrade	&	NC	&	US	&	NSTX-U	&	5	&	2001	&	2003	\\
Nuclear Science User Facilities	&	NC	&	US	&	NSUF	&	362	&	2006	&	2024	\\
Ocean Observatiories Initiative	&	NC	&	US	&	OOI	&	373	&	2003	&	2024	\\
The Molecular Foundry	&	NC	&	US	&	TMF	&	3,581	&	1996	&	2024	\\
IceCube Neutrino Observatory	&	NO	&	US	&	ICNO	&	210	&	2003	&	2024	\\
High-Flux Isotope Reactor	&	NS	&	US	&	HFIR	&	2,354	&	1993	&	2024	\\
Spallation Neutron Source	&	NS	&	US	&	SNS	&	2,803	&	2008	&	2024	\\
Accelerator Test Facility	&	PP	&	US	&	ATF	&	163	&	1990	&	2024	\\
Argonne Tandem Linac Accelerator System	&	PP	&	US	&	ATLAS	&	14	&	2015	&	2018	\\
Continuous Electron Beam Accelerator Facility	&	PP	&	US	&	CEBAF	&	3,208	&	1982	&	2024	\\
Fermilab Accelerator Complex	&	PP	&	US	&	FAC	&	149	&	2011	&	2024	\\
Facility for Advanced Accelerator Experimental Tests-II	&	PP	&	US	&	FACET-II	&	120	&	2014	&	2024	\\
Facility for Rare Isotope Beams	&	PP	&	US	&	FRIB	&	2,508	&	1991	&	2024	\\
Relativistic Heavy Ion Collider	&	PP	&	US	&	RHIC	&	35	&	2020	&	2024	\\
Advanced Light Source	&	SLS	&	US	&	ALS	&	17,007	&	1991	&	2024	\\
Advanced Photon Source	&	SLS	&	US	&	APS	&	33,523	&	1985	&	2024	\\
Cornell High Energy Synchrotron Source	&	SLS	&	US	&	CHESS	&	1,291	&	1997	&	2024	\\
Diamond	&	SLS	&	US	&	DIAMOND	&	12,042	&	1983	&	2024	\\
National Synchrotron Light Source	&	SLS	&	US	&	NSLS	&	7,505	&	1992	&	2024	\\
National Synchrotron Light Source-II	&	SLS	&	US	&	NSLS-II	&	3,115	&	2010	&	2024	\\
Stanford Synchrotron Radiation Lightsource	&	SLS	&	US	&	SSRL	&	13,572	&	1974	&	2024	\\

\hline
\hline
\end{longtable}
\end{spacing}

\clearpage
\begin{table}[htbp]
\caption{\textbf{Step wised regression of novelty papers}. Model (1) only includes the independent variable and three fixed-effects; Model (2) adds the paper-level confounding variables; Model (3) is estimated with additionally considering journal-level variables. Standard-errors are shown in parentheses and the significant Codes: ***: p$<$0.01, **: p$<$0.05, *: p$<$0.1.}
\label{tab:novel_step}
\begingroup
\centering
\footnotesize
\renewcommand{\arraystretch}{1}
\begingroup
\centering
\begin{tabular}{lccc}
\midrule \midrule
   Dependent Variable: & \multicolumn{3}{c}{Novel paper}\\
   Model:   & (1)  & (2)  & (3)\\
   \midrule
   BSF pub = True & 0.135$^{***}$   & 0.078$^{***}$  & 0.084$^{***}$ \\
                  & (0.006) & (0.006) & (0.006)\\
   \hline
   Number of authors (log)   && 0.248$^{***}$  & 0.257$^{***}$\\   
   && (0.005) & (0.005)\\
   Number of institutions (log)     && 0.009& 0.010\\   
   && (0.006) & (0.006)\\
   Number of countries (log)  && -0.281$^{***}$& -0.282$^{***}$\\    
   && (0.009) & (0.009)\\
   Number of references (log)&& 0.146$^{***}$ & 0.155$^{***}$\\     
   && (0.003) & (0.003)\\
   Leadership = all North  && 0.131$^{***}$  & 0.133$^{***}$\\  
   && (0.009) & (0.009)\\
   Leadership = all South  && -0.013        & -0.015 \\     
   && (0.014) & (0.014)\\
   Avg. career age (log)        && 0.031$^{***}$  & 0.032$^{***}$\\   
   && (0.005) & (0.005)\\
   Avg. institution h-index (log)&& -0.028$^{***}$ & -0.025$^{***}$\\   
   && (0.004) & (0.004)\\
   \hline
   Journal h-index (log) &&& -0.045$^{***}$\\   
   &&& (0.002)\\
   Core Journal = True    &&& 0.082$^{***}$\\ 
   &&& (0.012)\\
   \hline
   \emph{Fixed-effects}\\
   Discipline (26 fields) & Yes & Yes & Yes\\
   Last author & Yes & Yes & Yes\\  
   Publication year & Yes & Yes & Yes\\  
   \midrule
   \emph{Fit statistics}\\
   Observations  & 2,682,343      & 2,682,339      & 2,682,339\\ 
   Pseudo R$^2$  & 0.184       & 0.186        & 0.186\\
   \midrule \midrule
\end{tabular}
\par\endgroup
\par\endgroup
\end{table}

\begin{table}[htbp]
\caption{\textbf{Step wised regression of interdisciplinary papers.} Model (1) only includes the independent variable and three fixed-effects; Model (2) adds the paper-level confounding variables; Model (3) is estimated with additionally considering journal-level variables. Standard-errors are shown in parentheses and the significant Codes: ***: p$<$0.01, **: p$<$0.05, *: p$<$0.1.}
\label{tab:interdisciplinary_step}
\begingroup
\centering
\footnotesize
\renewcommand{\arraystretch}{1}
\begingroup
\centering
\begin{tabular}{lccc}
\midrule \midrule
   Dependent Variable: & \multicolumn{3}{c}{Interdisciplinary papers}\\
   Model:   & (1)  & (2)  & (3)\\  
   \midrule
   BSF pub = True    & 0.003$^{***}$ & 0.003$^{***}$ & 0.003$^{***}$\\   
   & (0.000) & (0.000) &(0.000)\\
   \hline
   Number of authors (log)         &      & -0.0001 & $5.38\times 10^{-5}$\\    
   && (0.000) & (0.000)\\
   Number of institutions (log) &      & 0.001$^{***}$ & 0.001$^{***}$ \\  
   && (0.000) & (0.000)\\
   Number of countries (log)        &      & -0.003$^{***}$ & -0.003$^{***}$\\   
   && (0.000) & (0.000)\\
   Number of references (log)      &      & 0.002$^{***}$ & 0.002$^{***}$\\   && (0.000) & (0.000)\\
   Leadership = all North  &      & 0.001$^{***}$ & 0.001$^{***}$\\   
   && (0.000) & (0.000)\\
   Leadership = all South  &      & 0.0008$^{*}$  & 0.0008$^{*}$\\   
   && (0.000) & (0.000)\\
   Avg. career age (log)   &      & 0.001$^{***}$ & 0.001$^{***}$\\   
   && (0.000) & (0.000)\\
   Avg. institution h-index (log)      &      & 0.0003$^{***}$& 0.0004$^{***}$\\
   && (0.000) & (0.000)\\
   \hline
   Journal h-index (log)    &      &     & -0.001$^{***}$\\  
   &&& (0.000)\\
   Core Journal = True         &      &     & $4.81\times 10^{-5}$\\ 
   &&& (0.000)\\
   \hline
   \emph{Fixed-effects}\\
   Discipline (26 fields) & Yes & Yes & Yes\\
   Last author & Yes & Yes & Yes\\  
   Publication year & Yes & Yes & Yes\\  
   \midrule
   \emph{Fit statistics}\\
   Observations   & 2,768,161      & 2,768,157     & 2,768,157\\  
   R$^2$  & 0.513       & 0.514       & 0.514\\
   \midrule \midrule
\end{tabular}
\par\endgroup
\par\endgroup
\end{table}

\begin{table}[htbp]
\caption{\textbf{Estimating the likelihood of producing novel papers across four research domains.} column (1) records the domain of physical sciences; column (2) records the domain of life sciences; column (3) shows the domain of health sciences and column (4) displays the domain of social sciences. Standard-errors are shown in parentheses and the significant Codes: ***: p$<$0.01, **: p$<$0.05, *: p$<$0.1.}
\label{tab:novel_by_domain}
\begingroup
\centering
\footnotesize
\renewcommand{\arraystretch}{0.9}
\begingroup
\centering
\begin{tabular}{lcccc}
\midrule \midrule
   Dependent Variable: & \multicolumn{4}{c}{Novel paper = True}\\
   Model:& (1)& (2)& (3)& (4)\\  
   Research domain: & Physical sciences & Life sciences & Health sciences & Social sciences \\
   \midrule
   BSF pub = True& 0.075$^{***}$  & 0.055$^{***}$  & 0.191$^{***}$  & 0.590$^{***}$\\   
 & (0.007)   & (0.012)   & (0.019)   & (0.091)\\   
   Number of authors (log)& 0.212$^{***}$  & 0.435$^{***}$  & 0.403$^{***}$  & 0.377$^{***}$\\
& (0.005)   & (0.009)   & (0.012)   & (0.054)\\   
   Number of institutions (log)& -0.004& 0.059$^{***}$  & 0.065$^{***}$  & -0.058\\   
& (0.007)   & (0.011)   & (0.015)   & (0.066)\\   
   Number of countries (log)& -0.293$^{***}$ & -0.210$^{***}$ & -0.196$^{***}$ & -0.383$^{***}$\\   
  & (0.010)   & (0.017)   & (0.024)   & (0.092)\\   
   Number of references (log)& 0.146$^{***}$  & 0.273$^{***}$  & 0.447$^{***}$  & 0.283$^{***}$\\   
   & (0.003)   & (0.006)   & (0.009)   & (0.030)\\   
   Leadership = all North   & 0.139$^{***}$  & 0.061$^{**}$   & 0.117$^{***}$  & 0.076\\   
  & (0.009)   & (0.021)   & (0.029)   & (0.093)\\   
   Leadership = all South   & -0.054$^{**}$  & -0.012& 0.083& -0.082\\   
  & (0.0015)   & (0.033)   & (0.046)   & (0.128)\\   
   Avg. career age (log) & 0.004& 0.123$^{***}$  & 0.108$^{***}$  & 0.006\\& (0.005)   & (0.010)   & (0.014)   & (0.055)\\   
   Avg. institution h-index (log) & -0.016$^{***}$ & -0.084$^{***}$ & -0.037$^{***}$ & 0.013\\
  & (0.004)   & (0.007)   & (0.010)   & (0.035)\\   
   Journal h-index (log) & -0.028$^{***}$ & -0.096$^{***}$ & -0.061$^{***}$ & 0.102$^{***}$\\  
& (0.002)   & (0.004)   & (0.006)   & (0.023)\\   
   Core Journal = True & 0.043$^{***}$  & 0.200$^{***}$  & 0.199$^{***}$  & 0.021\\   
  & (0.013)   & (0.030)   & (0.040)   & (0.103)\\   
   \midrule
   \emph{Fixed-effects}\\
   Discipline (26 fields) & Yes & Yes & Yes & Yes\\
   Last author & Yes & Yes & Yes & Yes\\  
   Publication year & Yes & Yes & Yes & Yes\\  
   \midrule
   \emph{Fit statistics}\\
   Observations & 2,236,396 & 683,985   & 337,939   & 26,000\\  
   Pseudo R$^2$ & 0.184   & 0.203   & 0.189   & 0.187\\  
   \midrule \midrule
   \multicolumn{5}{l}{\emph{Clustered (Last author) standard-errors in parentheses}}\\
   \multicolumn{5}{l}{\emph{Signif. Codes: ***: 0.01, **: 0.05, *: 0.1}}\\
\end{tabular}
\par\endgroup
\par\endgroup
\end{table}

\begin{table}[htbp]
\caption{\textbf{Estimating the level of producing interdisciplinary papers across four research domains. } column (1) records the domain of physical sciences; column (2) records the domain of life sciences; column (3) shows the domain of health sciences and column (4) displays the domain of social sciences. Standard-errors are shown in parentheses and the significant Codes: ***: p$<$0.01, **: p$<$0.05, *: p$<$0.1.}
\label{tab:interdisciplinary_by_domain}
\begingroup
\centering
\footnotesize
\renewcommand{\arraystretch}{0.9}
\begingroup
\centering
\begin{tabular}{lcccc}
\midrule \midrule
   Dependent Variable: & \multicolumn{4}{c}{Interdisciplinary papers}\\
   Model:& (1)& (2)& (3)& (4)\\  
   Research domain: & Physical sciences & Life sciences & Health sciences & Social sciences \\
   \midrule
   BSF pub = True & -0.001$^{***}$ & 0.014$^{***}$ & 0.019$^{***}$ & 0.007$^{***}$\\   
   & (0.000)& (0.000) & (0.000)& (0.001)\\   
   Number of authors (log)& $-7.42\times 10^{-5}$   & 0.001$^{***}$ & -0.0002& 0.005$^{***}$\\
    & (0.000)& (0.000)& (0.000)& (0.001)\\   
   Number of institutions (log)& 0.0008$^{***}$& 0.002$^{***}$ & 0.002$^{***}$& 0.0006\\   
  & (0.000)& (0.000)& (0.000)& (0.001)\\   
   Number of countries (log)  & -0.003$^{***}$& -0.001$^{***}$& 0.0001 & -0.004$^{**}$\\   
  & (0.000)& (0.000)& (0.000)& (0.002)\\   
   Number of references (log) & 0.003$^{***}$ & 0.002$^{***}$ & 0.002$^{***}$& 0.002$^{***}$\\   
  & (0.000)  & (0.000)& (0.000)& (0.001)\\   
   Leadership = all North   & 0.001$^{***}$ & 0.001$^{***}$ & 0.0004 & 0.003\\   
  & (0.000)& (0.000)& (0.000)& (0.002)\\   
   Leadership = all South   & 0.0007  & $-4.07\times 10^{-5}$   & 0.0005 & 0.0007\\  
  & (0.000)& (0.000)& (0.001)& (0.002)\\   
   Avg. career age (log)& 0.0005$^{***}$& 0.002$^{***}$ & $-1.13\times 10^{-5}$  & 0.002\\   
  & (0.000)& (0.000)& (0.000)& (0.001)\\   
   Avg. institution h-index (log) & -0.0003$^{***}$& 0.0005$^{***}$& 0.002$^{***}$& 0.001$^{*}$\\   
  & (0.000)& (0.000)& (0.000)& (0.001)\\   
   Journal h-index (log)& -0.001$^{***}$& $-4.59\times 10^{-6}$   & 0.0007$^{***}$& 0.0006\\   
  & (0.000)   & (0.000)  & (0.000)& (0.000)\\   
   Core Journal = True& $-1.81\times 10^{-5}$   & 0.0001  & 0.002$^{***}$& 0.005$^{**}$\\   
  & (0.000)& (0.000)& (0.001)& (0.002)\\   
   \midrule
   \emph{Fixed-effects}\\
   Discipline (26 fields) & Yes & Yes & Yes & Yes\\
   Last author & Yes & Yes & Yes & Yes\\  
   Publication year & Yes & Yes & Yes & Yes\\  
   \midrule
   \emph{Fit statistics}\\
   Observations & 2,321,151 & 729,686   & 376,452   & 41,437\\  
   Pseudo R$^2$ & 0.563   & 0.586   & 0.646   & 0.715\\  
   \midrule \midrule
   \multicolumn{5}{l}{\emph{Clustered (Last author) standard-errors in parentheses}}\\
   \multicolumn{5}{l}{\emph{Signif. Codes: ***: 0.01, **: 0.05, *: 0.1}}\\
\end{tabular}
\par\endgroup
\par\endgroup
\end{table}